\documentclass[a4paper]{article}   
\usepackage{epsbox}             
\usepackage{array}
\title{{\it ab initio} Study of Strain-Induced Ferroelectricity in SrTiO$_{3}$}
\author{Takatoshi~\textsc{Hashimoto}$^{1,2}$
\footnote{Corresponding author. E-mail address: hasitaka@imr.edu},
Takeshi~\textsc{Nishimatsu}$^{1}$,\\
Hiroshi~\textsc{Mizuseki}$^{1}$,
Yoshiyuki~\textsc{Kawazoe}$^{1}$,\\
Atsushi~\textsc{Sasaki}$^{2}$,
and
Yoshiaki~\textsc{Ikeda}$^{2}$\\
$^{1}$Institute for Materials Research (IMR),\\Tohoku University, Sendai 980-8577, Japan\\
$^{2}$NEC TOKIN Corporation, Sendai 982-8510, Japan}
\begin{document}
\maketitle
\sloppy
\section*{Abstract}
Valley lines on total-energy surfaces for the zone-center distortions
of free-standing and in-plane strained SrTiO$_{3}$ are investigated
with a newly developed first-principles structure optimization technique
[Jpn. J. Appl. Phys. {\bf 43} (2004), 6785].
The results of numerical calculations confirmed
that the ferroelectricity is induced,
and the Curie temperature is increased,
by applying biaxial compressive or tensile strains.
Along the distortion, strong nonlinear coupling between the 
soft- and hard-modes is demonstrated.
\subsection*{Keywords}
density functional theory, local density approximation,
biaxial strain, potential surface, Slater mode, Last mode
\section{Introduction}
\begin{figure}
  \begin{center}
    \epsfile{file=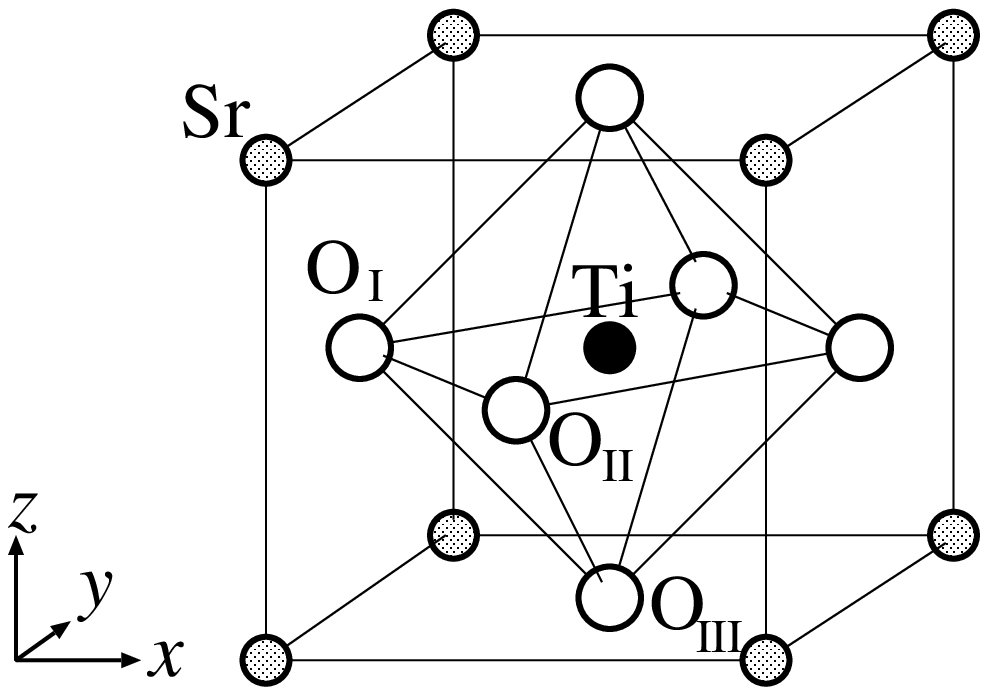,width=4cm}
  \end{center}
  \caption{Cubic ($O_{h}^{1}$) crystal structure of perovskite oxide SrTiO$_3$.}
  \label{fig:structure}
\end{figure}

Perovskite structure strontium titanate, SrTiO$_{3}$,
is an extremely important material.  Because of its wide
range of physical properties, such as semiconductivity,
superconductivity, incipient ferroelectricity and catalytic activity,
SrTiO$_{3}$ is expected to be used for
a variety of technological applications.
Bulk SrTiO$_{3}$ crystallizes in the cubic 
centrosymmetric ($O_{h}^{1}$) structure,
and is paraelectric at room temperature.
Atomic fractional coordinates are
Sr(0,0,0),
Ti($\frac{1}{2},~\frac{1}{2},~\frac{1}{2}$),
O$_{\rm I}$($0,~\frac{1}{2},~\frac{1}{2}$),
O$_{\rm II}$($\frac{1}{2},~0,~\frac{1}{2}$), and
O$_{\rm III}$($\frac{1}{2},~\frac{1}{2},~0$)
respectively,
as depicted in Fig.~\ref{fig:structure}.
SrTiO$_{3}$'s tolerance factor~\cite{Goldschmidt:ToleranceFactor},
calculated with Shannons's ionic radii~\cite{ShannonEffectiveIonicRadii}, is $t=1.002$.
This almost unity value suggests that the cubic structure is moderately stable.
Below 105K, SrTiO$_{3}$ undergoes a structural phase transition
from cubic to an antiferrodistortive (AFD) tetragonal ($D_{4h}^{18}$) structure,
which is associated with zone-boundary phonon condensation
at the R~point.~\cite{Landolt:Bornstein:SrTiO3,
Unoki:S:JPhysSocJpn:23:p546-552:1967,
Fleury:S:W:PRL:21:p16:1968,
Shirane:Y:PhysRev:177:p858:1969,
OKAI:Y:JPhysSocJpn:39:p162-165:1975}
This transition is driven by TiO$_{6}$ octahedral rotational mode instabilities;
the rotation angle is less than 2$^{\circ}$~\cite{Muller:B:PRL:26:p13:1970}.
At lower temperatures, the dielectric constant
exhibits Curie-Weiss law like behavior, but does not exhibit divergence.
The dielectric constant saturates to a value of $\sim 2\!\times\!10^4$ under
10~K~\cite{VIANA:L:H:B:L:PRB:50:p601-604:1994,Itoh:W:I:Y:S:N:PRL:82:p3540-3543:1999},
but no transition, e.g. to a ferroelectric state, actually occurs with decreasing 
temperature. Because of its failure to exhibit a ferroelectric phase transition,
SrTiO$_{3}$ is generally regarded as an incipient ferroelectric, in which
quantum fluctuations suppress ferroelectricity.
The observed temperature dependence of SrTiO$_{3}$'s dielectric constant
is well described by Barrett's formula\cite{Barrett:PhysRev:86:p118-120:1952},
\begin{equation}
  \label{eq:BarrettFormula}
  \varepsilon = \frac{M}{\frac{1}{2}T_1\coth\Big(\frac{T_1}{2T}\Big)-T_0} ,
\end{equation}
where $T_0$ is the {\it transition} temperature,
$T_1$ is a characteristic dividing point below which quantum effects are important,
and $M$ is a constant.
In ref.~\cite{Sawaguchi:K:K:JPhysSocJpn:18:p459:1963},
these values for SrTiO$_{3}$ were given as
$T_0=38~\rm K$, $T_1=84~\rm K$, and $M=9\times 10^4~\rm K$, respectively.
Thus, it had been long expected that
SrTiO$_{3}$ would have ferroelectricity under some stresses or strains.

In recent years,
the preparation of oxide thin films
by thermal non-equilibrium techniques such as
molecular beam epitaxy (MBE) and
pulsed laser deposition (PLD),
have enabled the growth of hetero-epitaxial thin films,
and have attracted a great deal of attention.
In 2004,
J. H. Haeni {\it et al.} demonstrated that thin films of SrTiO$_{3}$
are ferroelectric near room 
temperature.~\cite{Haeni:I:C:U:R:L:C:T:H:C:T:P:S:C:K:L:S:Nature:430:p758-761:2004}
Using MBE, they grew SrTiO$_{3}$ thin films
on a DyScO$_{3}$ substrate, to
induce a biaxial tensile strain of the order of 1\%,
and measured the temperature dependence of the
in-plane dielectric constant.
Their results indicate that the nonpolar ground state of SrTiO$_{3}$
can be drastically transformed, to a polar state, by the
application of small strains.

{\it Ab initio} calculations are attractive 
for atomistically analyzing the effects of 
structural distortions
on properties such as Curie temperature.
T. Schimizu studied
the frequencies of $\Gamma$ transverse-optical phonons
in the cubic SrTiO$_{3}$
using the frozen-phonon scheme,
and found that
the dielectric constant strongly depends on strain
at finite temperature~\cite{Schimizu:1997}.
A. Antons {\it et al.} have confirmed
that the dielectric constant of SrTiO$_{3}$ epitaxial thin films
varies significantly with strain
using density functional theory (DFT)
within the local-density approximation (LDA)~\cite{Antons:N:R:V:PRB:71:p024102:2005}.
However, the mechanism by which in-plane biaxial strain induces
ferroelectricity in SrTiO$_{3}$ has not been clarified.
We investigate the relationship between 
SrTiO$_{3}$-ferroelectricity and in-plane biaxial strains, compressive and tensile, 
by analyzing valley lines of total-energy surfaces
with {\it ab initio} structure optimization technique
that we had developed~\cite{Hashimoto:N:M:K:S:I:JJAP:43:p6785-6792:2004}.
Compared to the conventional soft-mode-only structure optimization technique,
our approach has the advantages
that it can: (1) accurately estimate total energy
as a function of the amplitude of atomic displacements,
(2) correctly investigate nonlinear coupling
between soft-mode atomic displacements,
hard-mode displacements, and lattice deformations.

In the next section,
we briefly explain our structure optimization technique
and the numerical methods used in this study.
Results of calculations are shown in \S\ref{sec:results}.
In \S\ref{sec:summary}, we summarize the paper.

\section{Method}
\label{sec:TechnicalDetailsOfCalculations}

\subsection{structure optimization technique}
\label{subsec:Formalism}
\begin{figure}
  \begin{center}
    \epsfile{file=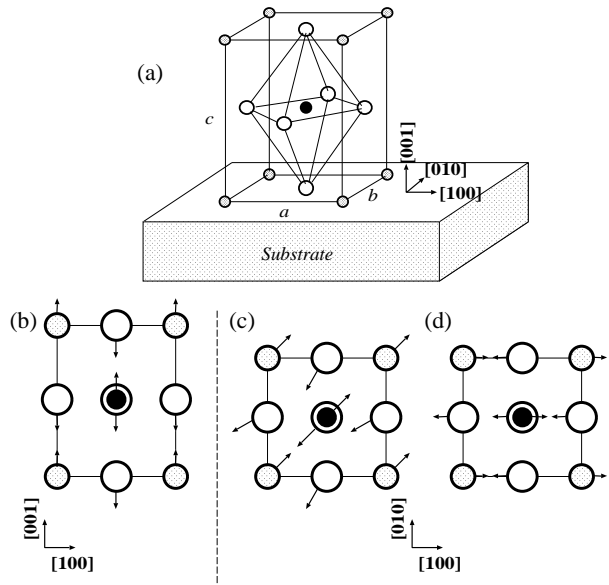,width=8.0cm}
  \end{center}
  \caption{(a) Exaggerated illustration of epitaxially grown SrTiO$_{3}$ on substrate.
           Lattice constants $a$, $b(=a)$, and $c$ and crystallographic directions
           [100], [010], and [001] are indicated.
           (b) Pattern of atomic displacements accompanying
           [001] tetragonal distortion under biaxial compressive strain
           is indicated by arrows in $(0\bar{1}0)$ projection.
           (c) That of [110] monoclinic distortion under biaxial tensile strain
           in $(100)$ projection.
           (d) That of [100] monoclinic distortion under biaxial tensile strain
           in $(100)$ projection.}
  \label{fig:DistortionsOfSrTiO3}
\end{figure}

To investigate the total-energy surfaces of perovskite oxides,
we improve King-Smith and Vanderbilt's scheme~\cite{King-Smith:V:1994}
and redefine
the amplitude of atomic displacements $u_{\alpha}$ as
\begin{equation}
  \label{eq:redefinition}
     u_{\alpha}=\sqrt{
                 \left( v_\alpha^A         \right)^2
                +\left( v_\alpha^B         \right)^2
                +\left( v_\alpha^{\rm O_I} \right)^2
                +\left( v_\alpha^{\rm O_{II}} \right)^2
                +\left( v_\alpha^{\rm O_{III}} \right)^2}\ ,
\end{equation}
where $v_{\alpha}^{\tau}$ is the displacement
of each atom $\tau$ ($= A, B,
{\rm O_I},
{\rm O_{II}},
{\rm O_{III}}$)
in the Cartesian directions of $\alpha ($=x, y, z$)$
from the symmetric perovskite structure
($O_{h}^1$ or $D_{4h}^1$).
The total energy is evaluated
as a function of $u=\! \sqrt{u_{x}^{2}+u_{y}^{2}+u_{z}^{2}}~$
under the condition that
$v_{\alpha}^{\tau}$
and the strain components $\eta_i$~($i=1,\ \dots,\ 6$; Voigt notation)
minimize the total energy for each $u$
using an {\it ab initio} norm-conserving pseudopotential method
and geometric optimization.
To simulate $D_{4h}^1$ epitaxial SrTiO$_{3}$ thin films,
as illustrated in Fig.~\ref{fig:DistortionsOfSrTiO3}(a),
under a substrate induced biaxial in-plane strain, 
only the $c$ axis
(i.e., $\eta_{3}$) is permitted to relax
but other parameters $a=b$, $\alpha=\beta=\gamma=90^{\circ}$
(i.e., $\eta_{1}, \eta_{2}, \eta_{4}, \eta_{5},$ and $\eta_{6}$)
are fixed for all directions of polarizations,
[001] (Fig.~\ref{fig:DistortionsOfSrTiO3}(b)),
[110] (Fig.~\ref{fig:DistortionsOfSrTiO3}(c)),
and [100] (Fig.~\ref{fig:DistortionsOfSrTiO3}(d)),
to satisfy symmetry constraints.
Total energy minimization
under the constant-$u_{\alpha}$ constraint
(i.e., on the sphere surface with a radius $u_{\alpha}$)
is performed iteratively.
Fully detailed formalism of this structure optimization technique
is given in ref.~\cite{Hashimoto:N:M:K:S:I:JJAP:43:p6785-6792:2004}.
We continue iterative
optimization of atomic and lattice structure
until differences in total energies remain less than
$10^{-7}$ Hartree for two successive iterations.

\subsection{Calculation Methods}
For all {\it ab initio} calculations,
we use the ABINIT package~\cite{Gonze:ABINIT.ComputMaterSci:2002}
with adapting it for our structure optimization technique
in its source code of Src\_9drive/brdmin.f.
Bloch wave functions of electrons are expanded into
plane waves with a energy cutoff of 40~Hartree,
using Teter's extended norm-conserving pseudopotentials.~\cite{Teter:Pseudopotential:1993}
To retain a constant energy cutoff
and to avoid nonphysical discontinuities of total energies as function of volumes,
a correction of kinetic energies of plane waves
just below the energy cutoff is introduced~\cite{BERNASCONI:C:F:S:T:P:1995}
as implemented in the ABINIT.
This ``energy cutoff smearing'' technique
is efficient not only for constant-pressure molecular dynamics
but also for comparing total energies in different volumes.
With this technique,
one does not have to achieve convergence
between constant-number-of-plane-waves calculations
and constant-energy-cutoff calculations
with respect to the basis set
and, therefore, does not have to adopt
a superfluously large energy cutoff.
The pseudopotentials include
O 2s and 2p,
Ti 3s, 3p, 3d and 4s,
Sr 4s, 4p and 5s,
as valence electrons.
Bloch wave functions are sampled on an
$8\!\times\! 8\!\times\! 8$ grid of $k$-points
in the first Brillouin zone.
The exchange-correlation energy is treated
within the local density approximation (LDA).
As the parametrized correlation energy,
we use Teter's rational polynomial parameterization,~\cite{Goedecker:T:H:1996}
which reproduces the results
obtained by Ceperley and Alder.~\cite{CeperleyAlder}
The electronic states are calculated by the iterative scheme
to reach a tolerance of convergence that requires
differences of forces to be less
than $5\!\times\! 10^{-7}$ Hartree/Bohr for two successive iterations.

With these calculation methods,
the calculated equilibrium lattice constant
for cubic SrTiO$_{3}$ is 7.27~Bohr,
which is $\sim 1.3\%$ less than the experimental value
of 7.36~Bohr
obtained by linearly extrapolating the lattice constant
at high temperature to zero temperature
(see Fig.~1 in ref.~\cite{Okazaki:K:MaterResBull:8:p545-550:1973}).
This underestimation is commonly considered
to be a result of errors from LDA.
Properties of ferroelectrics, especially total-energy surfaces,
are very sensitive to the lattice constant~\cite{Cohen:Nature:1992}.
Semi-empirical constraints on lattice constants
in calculations of perovskite oxides have been commonly used
to acquire agreements between
the experimentally observed values
and calculated results.
Although artificial constraints can be introduced
in our structure optimization technique,
we do not employ such semi-empirical constraints.
Nevertheless, we believe that calculations using LDA
clarify some trends of displacive transitions in strained SrTiO$_{3}$.

The Berry-phase method~\cite{King-Smith:V:PRB:47:p1651-1654:1993,Resta:1994}
is used to evaluate spontaneous polarizations.

For simplicity and to emphasize the origin of ferroelectricity in
strained SrTiO$_{3}$,
we neglect the AFD instabilities 
and use a single unit cell to calculate total-energy surfaces.
We believe that this is a reasonable approximation
because the effect of AFD on dielectric response
is negligibly small.~\cite{Sawaguchi:K:K:JPhysSocJpn:17:p1666:1962}

\section{Results and Discussion}
\label{sec:results}
\subsection{Free-standing SrTiO$_{3}$}
\begin{figure}
  \begin{center}
    \epsfile{file=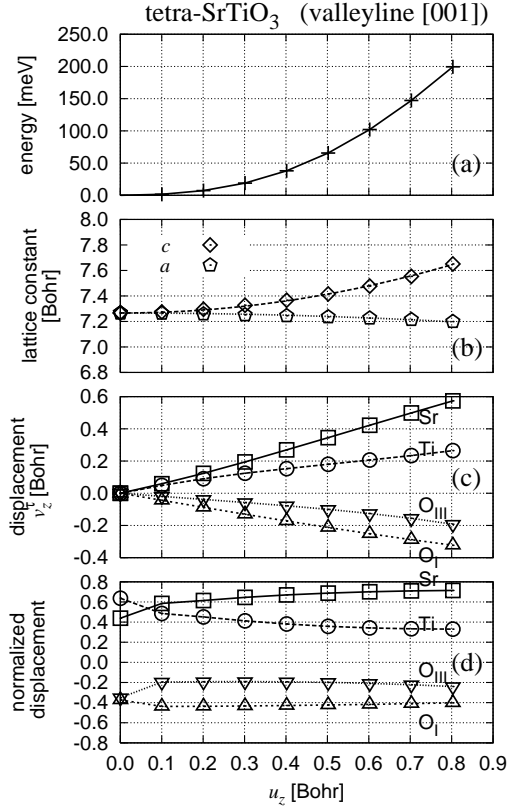,width=7.5cm}
  \end{center}
  \caption{Calculated results for free-standing SrTiO$_{3}$
   regarding atomic displacements of tetragonal [001] direction.
   (a)~Calculated total energy (in meV)
    as a function of $u_z$ (in Bohr) ($+$'s connected with solid lines).
    Zero of the energy scale is placed at
    the total energy of the cubic structure when $u_z=0$.
    (b)~Lattice constants $a$ and $c$ in Bohr as functions of $u_z$
    fitted by quadratic functions
    drawn with a dotted line and a dashed line, respectively.
    (c)~Atomic displacements $v_z^{\rm Sr}, v_z^{\rm Ti}, v_z^{\rm O_I}=v_z^{\rm O_{II}},
    v_z^{\rm O_{III}}$ as functions of $u_z$.
    (d)~Normalized atomic displacements $v_z^\tau/u_z$ as functions of $u_z$.
    The $\Gamma_{15}$ soft-mode eigenvector calculated by the frozen phonon method
    is additionally shown at $u_z=0$.}
  \label{fig:TotalEnergyFreeStandSrTiO3}
\end{figure}

Before examining in-plain strained SrTiO$_{3}$,
we evaluated the total-energy surface of
free-standing SrTiO$_{3}$ that is distorted from cubic
to tetragonal and has [001] polarization. 
Shown in Fig.~\ref{fig:TotalEnergyFreeStandSrTiO3}(a)
is the total energy as a function of the amplitude of atomic displacements $u_{z}$.
M.~Itoh {\it et al.} found that ferroelectricity was induced in SrTiO$_3$
by the isotope exchange of $^{18}$O for $^{16}$O.~\cite{Itoh:W:I:Y:S:N:PRL:82:p3540-3543:1999}
This isotope exchange experiment and
the incipient ferroelectricity with $T_0=38~{\rm K}>0$ in Barrett's formula
suggest that the total energy {\it vs} $u_{z}$ curve might have double well structure,
but our calculated result does not.
This is not just an artifact of the sensitivity of LDA total-energy surfaces
to lattice constants,
as descried in refs.~\cite{Cohen:Nature:1992}
and \cite{Sai:V:PRB:62:p13942-13950:2000},
rater, as we show in next subsection \S\ref{subsec:strain},
the system is {\it very close} to having a double well.

Calculated lattice constants $a$ and $c$ are well fitted
by quadratic functions in Fig.~\ref{fig:TotalEnergyFreeStandSrTiO3}(b).
It is surprising that the amplitude of Sr-displacement, $v_z^{\rm Sr}$,
is smaller than that of Ti, $v_z^{\rm Ti}$,
for the soft-mode eigenvector calculated by the frozen phonon method at $u_z=0$,
but $v_z^{\rm Sr}$ becomes larger than $v_z^{\rm Ti}$ for $u_z > 0.07$
as shown in Fig.~\ref{fig:TotalEnergyFreeStandSrTiO3}(c) and (d).

\subsection{Effects of in-plane strains on SrTiO$_{3}$}
\label{subsec:strain}
\begin{figure}
  \begin{center}
    \epsfile{file=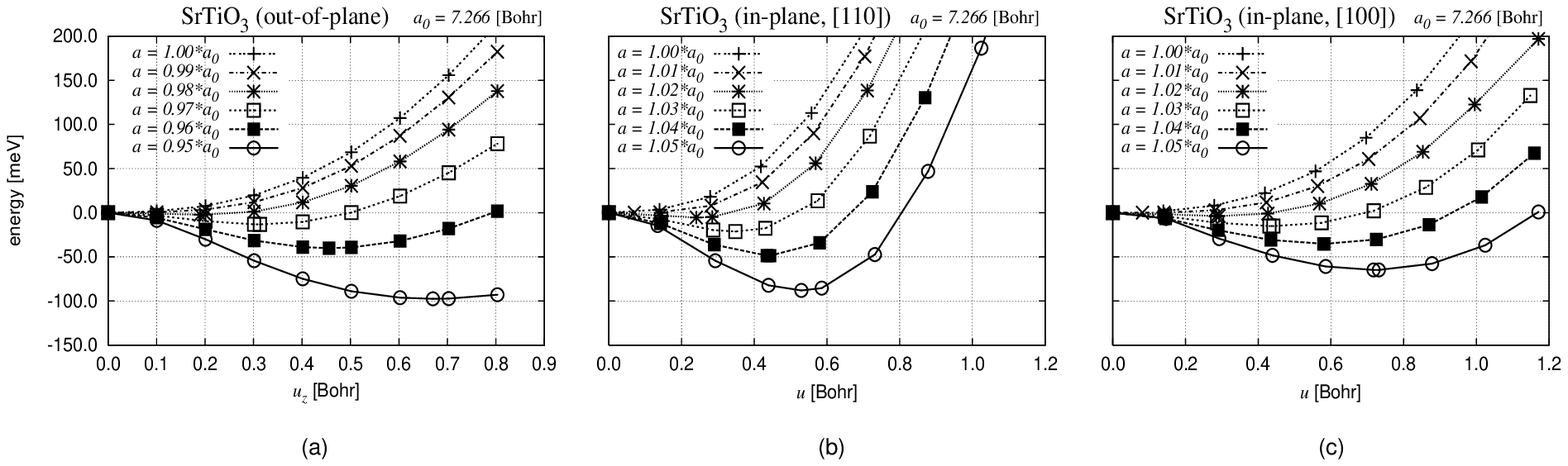,width=12cm}
  \end{center}
  \caption{Calculated total energies (in meV) of SrTiO$_{3}$
           as functions of $u=\sqrt{u_x^2+u_y^2+u_z^2}$ (in Bohr) for
           the tetragonal [001] distortion under in-plane biaxial compressive strains (a) and
           the monoclinic [110] (b) and [100] (c) distortions
           under in-plane biaxial tensile strains.
           Zero of the energy scale is placed
           at the total energy of the paraelectric structure (i.e., $u=0$) for each strain.}
  \label{fig:TotalEnergyCompare}
\end{figure}

\begin{figure}
  \begin{center}
    \epsfile{file=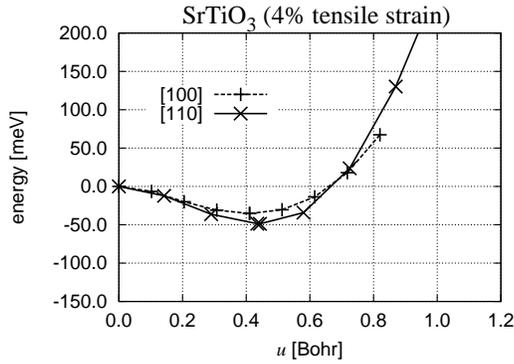,width=7.5cm}
  \end{center}
  \caption{Comparison of $u$-dependence of total energies (in meV)
           for the monoclinic [110] and [100] ferroelectric distortions
           under same biaxial tensile strain of 4\%.}
  \label{fig:TotalEnergyCompareMonoSrTiO3}
\end{figure}

\begin{figure}
  \begin{center}
    \epsfile{file=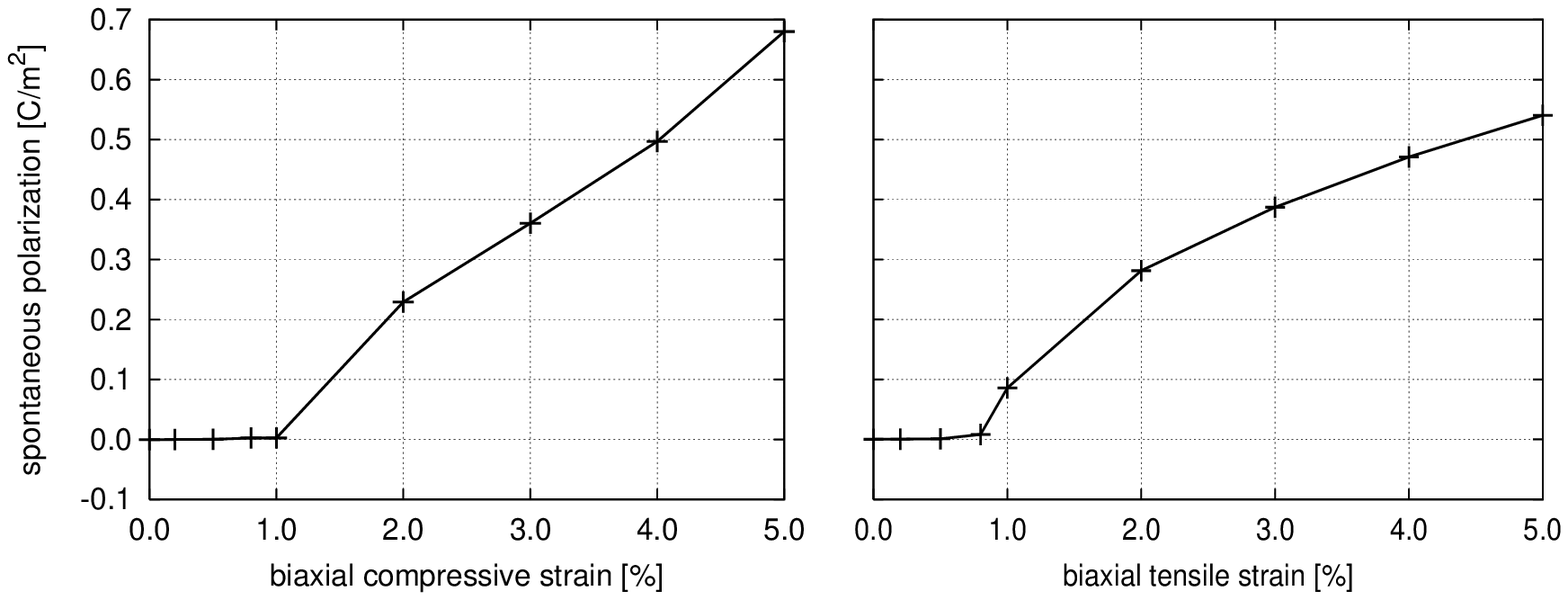,width=12cm}
  \end{center}
  \caption{Calculated spontaneous polarizations
           of the equilibrium structures
           as a function of applied biaxial strains.}
  \label{fig:polarSrTiO3}
\end{figure}

Total energies of SrTiO$_{3}$
as functions of the amplitude of atomic displacements $u$
are calculated
for the tetragonal [001] distortion
under in-plane biaxial compressive
strains (Fig.~\ref{fig:TotalEnergyCompare}(a)) and
the monoclinic [110] (Fig.~\ref{fig:TotalEnergyCompare}(b)) 
and [100] (Fig.~\ref{fig:TotalEnergyCompare}(c)) distortions
under in-plane biaxial tensile strains.
It is confirmed that, for the biaxial tensile strained SrTiO$_3$,
the [110] direction of ferroelectric polar distortion is energetically
more preferable than that of [100]
as compared in Fig.~\ref{fig:TotalEnergyCompareMonoSrTiO3}.
Energy gain,
the deference between total energy at $u=0$ and its minimum value,
grows up according to both compressive and tensile biaxial strains,
though it is zero in the zero-strain $a=a_0$ case.
The value of $u$,
at which the system exhibits minimum total energy and equilibrium structure,
also grows up according to the biaxial strains.
In Fig.~\ref{fig:polarSrTiO3}, we show
calculated spontaneous polarizations
for the equilibrium structures under biaxial strains $a=0.95 a_0 \sim 1.05 a_0$.
Under the compressive strain,
spontaneous polarization appears above 1\%
and increases linearly as a function of strain.
Under the tensile strain,
spontaneous polarization appears a little below 1\% and 
increases more gradually with strain.
These calculated results represent that
the transition temperature and the spontaneous polarization for
ferroelectric ordering are predicted to be monotonically
increasing functions of biaxial in-plane strains.
In our calculated results,
the energy gain becomes barely the order of room temperature (300\,K $\approx$ 26\,meV)
at 4\% of compressive or tensile strains.
The 4\% of compressive or tensile strains are
impractical even by the thermal non-equilibrium thin-film growth techniques,
and larger than the 1\% of tensile strain under which ferroelectricity
was experimentally found,~\cite{Haeni:I:C:U:R:L:C:T:H:C:T:P:S:C:K:L:S:Nature:430:p758-761:2004}
though it is difficult to compare experiments and LDA-based calculations.
On the other hand, Haeni {\it et al.} could not observe the ferroelectricity
in the 0.9\% compressive strained SrTiO$_{3}$ thin film on a (La,Sr)(Al,Ta)O$_{3}$
substrate.~\cite{Haeni:I:C:U:R:L:C:T:H:C:T:P:S:C:K:L:S:Nature:430:p758-761:2004}
It is also difficult to conclude
that 0.9\% compressive strain is less than the critical strain required 
to induce ferroelectricity
or that 0.9\% is enough to induce the ferroelectricity,
but a surface effect (or any other effects) suppresses it.

\begin{figure}
  \begin{center}
    \epsfile{file=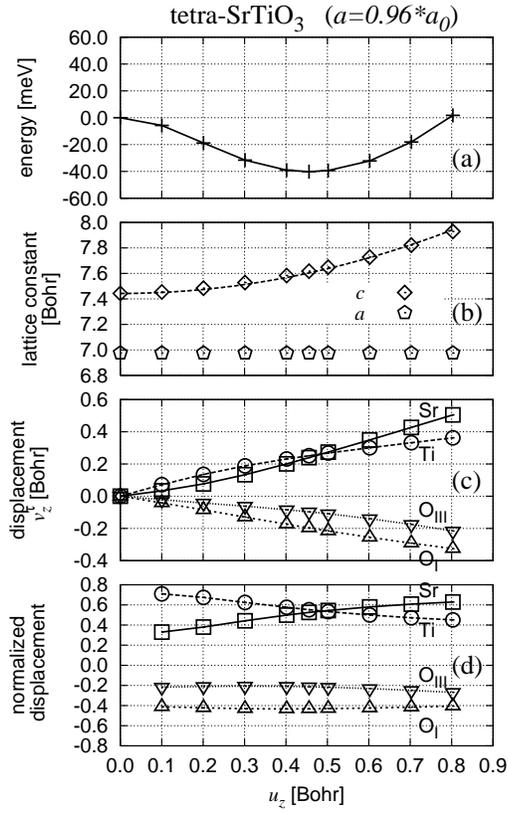,width=7.5cm}
  \end{center}
  \caption{Calculated results for SrTiO$_{3}$ under 4\% biaxial compressive strain
    regarding atomic displacements of tetragonal [001] direction.
    (a)~Calculated total energy (in meV)
    as a function of $u_z$ (in Bohr) ($+$'s connected with solid lines).
    Zero of the energy scale is placed at
    the total energy of the cubic structure when $u_z=0$.
    (b)~Lattice constants in Bohr. $a$ is fixed.
    $c$ is fitted by quadratic functions of $u_z$
    and drawn with a dashed line.
    (c)~Atomic displacements $v_z^{\rm Sr}, v_z^{\rm Ti}, v_z^{\rm O_I}=v_z^{\rm O_{II}},
    v_z^{\rm O_{III}}$ as functions of $u_z$.
    (d)~Normalized atomic displacements $v_z^\tau/u_z$ as functions of $u_z$.}
  \label{fig:TotalEnergyTetraSrTiO3}
\end{figure}

\begin{figure}
  \begin{center}
    \epsfile{file=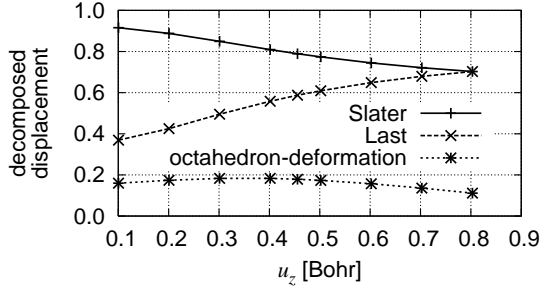,width=7.5cm}
  \end{center}
  \caption{Normalized atomic displacements
           of tetragonal SrTiO$_3$ are decomposed into
           three conventional modes:
           Slater mode (solid line),
           Last mode (dashed line),
           and the octahedron-deformation mode (dotted line)
           as functions of $u_z$ in Bohr.}
  \label{fig:SrTiO3TetraSlater}
\end{figure}

To clarify the structural effects of in-plane biaxial strain,
we analyze the detailed behavior of atomic displacements for 4\%
biaxial compressive and tensile strains:

Predicted results for
the compressive strain, as functions of $u_{z}$, are plotted in 
Fig.~\ref{fig:TotalEnergyTetraSrTiO3}:
(a) total energy; (b) lattice constants; (c) and (d) atomic displacements.
Ferroelectric tetragonal equilibrium, with a 40.2\,meV stabilization 
energy gain, is predicted at $u_{z}=0.456$\,Bohr.
The lattice constant $c(u_{z})$, is well fitted with a quadratic function. 
Normalized atomic displacements $v_z^{\tau}/u_z$
change greatly in the investigated range of $u_z$,
in contrast with BaTiO$_{3}$
in which they remain almost constant to
the $\Gamma_{15}$ soft-mode eigenvector
in ferroelectric distortion~\cite{Hashimoto:N:M:K:S:I:JJAP:43:p6785-6792:2004}.
In the atomic displacements of the positive charge,
we can find
that the displacement of Ti is dominant
from paraelectric $u_{z}=0$ state
to $u_{z}=0.456$ at which the total energy becomes minimum,
whereas the displacement of Sr become activated instead of Ti
in the region of $u>0.456$,
as shown in Fig.\ref{fig:TotalEnergyTetraSrTiO3}(c).
There may be a relatedness between this Sr-Ti-displacements-crossing point
and the ferroelectric equilibrium structure with the minimum total energy,
thus we mentioned that free-standing SrTiO$_3$, which have the crossing point
at around $u=0.07$, is {\it very close} to having a double well.
In the negative charge,
the atomic displacement of O$_{\rm I}$($\equiv$ O$_{\rm II}$)
is dominant compared to that of O$_{\rm III}$
throughout the investigated range of $u_{z}$.
To obtain the deeper insights about these behavior of the atomic displacements,
we analyze the presently obtained normalized atomic displacements
by decomposing them into conventional three
$\Gamma_{15}$ modes:
Slater mode,~\cite{Slater:1950}
Last mode,~\cite{Last:1957}
and the octahedron-deformation mode,
as listed in Table~\ref{tab:modes}.
As shown in Fig.~\ref{fig:SrTiO3TetraSlater},
although the Slater mode is dominant
through out the entire investigated range of atomic displacements,
the Last mode tend to increase while the Slater mode decreases moderately
as functions of the amplitude of atomic displacements $u_{z}$.
This result is corresponding to the sight
pointed out by Schimizu~\cite{Schimizu:1997}
and Harada~\cite{HARADA:A:S:ActaCrystallogr:A26:p608:1970}
that the Last mode is also important
while the Slater mode plays a key role
in the phase transition of SrTiO$_{3}$.
\begin{table}
  \caption{Conventional atomic displacive modes. Slater mode, Last mode,
    and the octahedron-deformation mode.
    The translational mode and the $\Gamma_{25}$ mode are also listed.}
  \label{tab:modes}
\begin{center}
\begin{tabular}{@{\hspace{\tabcolsep}%
      \extracolsep{\fill}}l|rrr|rr} \hline
     &               &               & octa.        &             &      \\
     & Slater        & Last          & deform.      & trans.      & $\Gamma_{25}$\\
     & mode          & mode          & mode         & mode        & mode \\
\hline
$A$  & 0~~~          & $ 4/\sqrt{20}$& 0~~~         &$ 1/\sqrt{5}$& 0~~~\\
$B$  & $ 3/\sqrt{12}$& $-1/\sqrt{20}$& 0~~~         &$ 1/\sqrt{5}$& 0~~~\\
O$_{\rm I}$& $-1/\sqrt{12}$& $-1/\sqrt{20}$& $ 1/\sqrt{6}$&$ 1/\sqrt{5}$& $ 1/\sqrt{2}$\\
O$_{\rm II}$& $-1/\sqrt{12}$& $-1/\sqrt{20}$& $ 1/\sqrt{6}$&$ 1/\sqrt{5}$& $-1/\sqrt{2}$\\
O$_{\rm III}$& $-1/\sqrt{12}$& $-1/\sqrt{20}$& $-2/\sqrt{6}$&$ 1/\sqrt{5}$& 0~~~ \\
\hline
\end{tabular}
\end{center}
\end{table}

\begin{figure}
  \begin{center}
    \epsfile{file=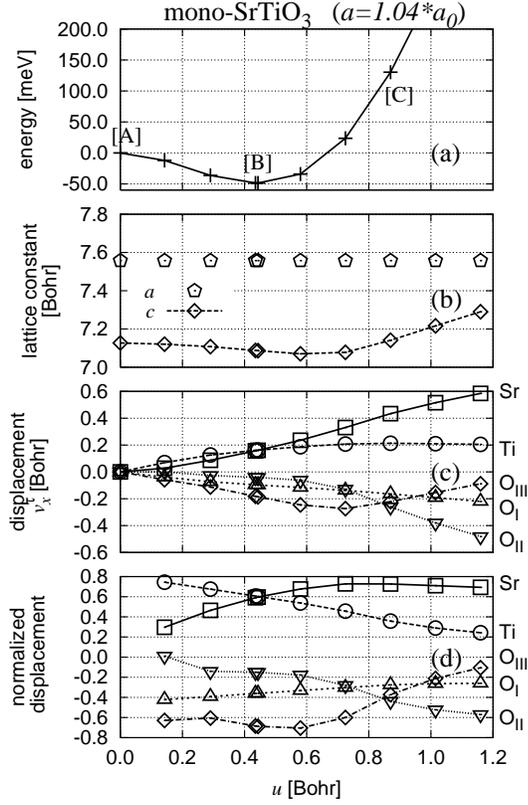,width=7.5cm}
  \end{center}
 \caption{Calculated results for SrTiO$_{3}$ under 4\% biaxial tensile strain
    regarding atomic displacements of tetragonal [110] direction.
    (a)~Total energy (in meV)
    as a function of $u$ (in Bohr) ($+$'s connected with solid lines).
    Zero of the energy scale is placed at
    the total energy of the cubic structure when $u=0$.
    (b)~Lattice constants $c$ and fixed $a$ in Bohr.
    (c)~Atomic displacements $v_x^{\tau}$.
    (d)~Normalized atomic displacements $v_x^{\tau}/u_x$ as functions of $u$.}
  \label{fig:TotalEnergyMonoSrTiO3}
\end{figure}

\begin{figure}
  \begin{center}
    \epsfile{file=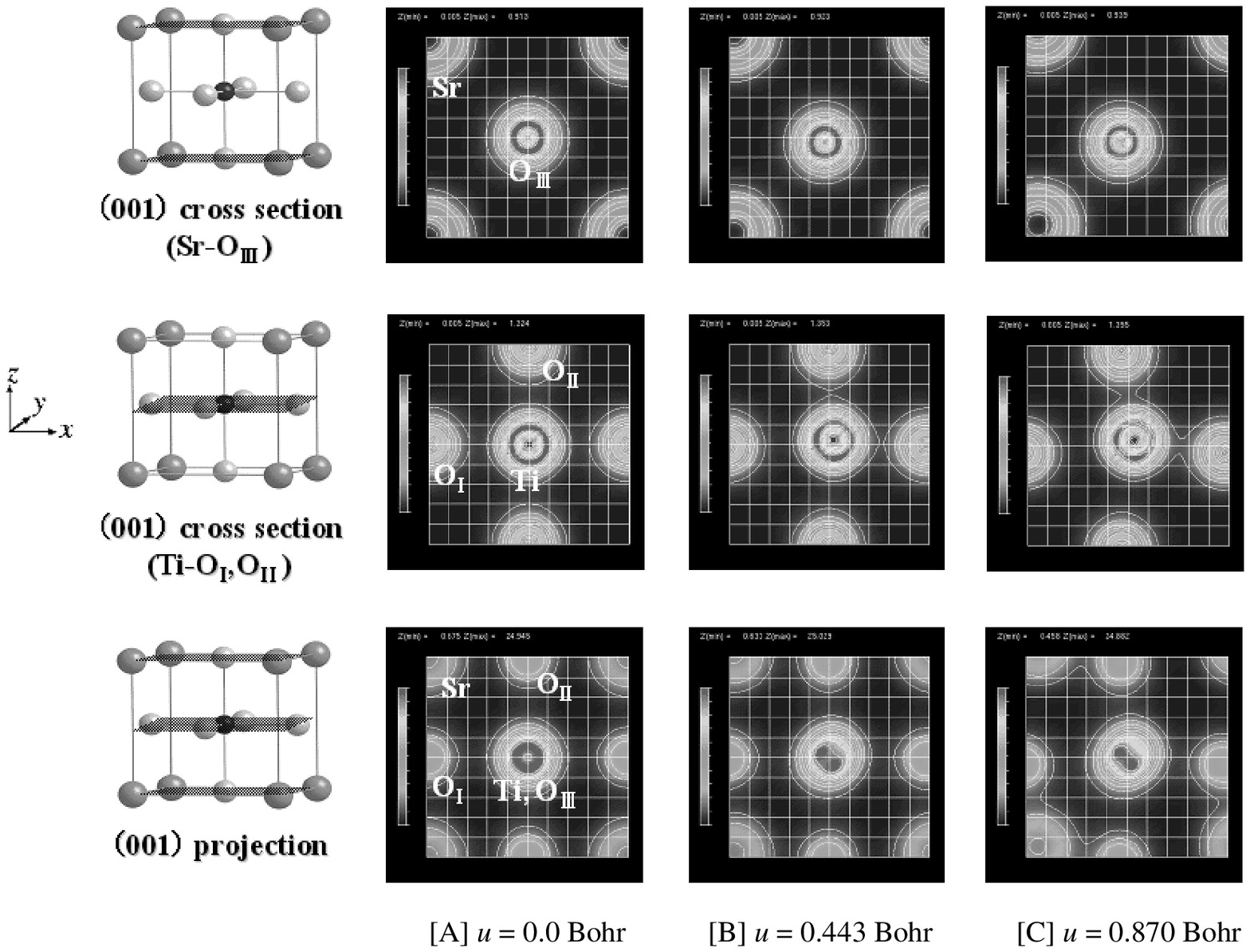,width=12cm}
  \end{center}
  \caption{Pseudo valence charge density maps
    in the (001) cross sections and a projection
    for SrTiO$_{3}$ under 4\% biaxial tensile strain
    at the following three points of atomic distortion;
    [A] paraelectric state at $u=0$,
    [B] monoclinic equilibrium state at $u=0.443$ Bohr, and
    [C] large displacement state at $u=0.870$  Bohr,
    which are indicated on total-energy surface
    shown in Fig.\ref{fig:TotalEnergyMonoSrTiO3}(a).
    Maps were drawn with VENUS.~\cite{Izumi:JCrystallogrSocJpn:2002}}
  \label{fig:ChargeMonoSrTiO3}
\end{figure}

For SrTiO$_{3}$ under the 4\% tensile strain,
the calculated results
of total energy, lattice constants and atomic displacements of $x$-direction
as functions of the amplitude of atomic displacements $u=\sqrt{u_x^{2}+u_y^{2}}$
are shown in Fig.~\ref{fig:TotalEnergyMonoSrTiO3}.
In this case,
ferroelectric monoclinic equilibrium structure is produced
by getting the 48.6\,meV energy gain at $u=0.443$\,Bohr.
In the atomic displacements of the $x$-direction,
Ti in the positive charge and O$_{\rm I}$
($\equiv$ O$_{\rm II}$ of $y$-direction)
and O$_{\rm III}$
in the negative charge
are dominant from paraelectric $u=0$ state
to $u=0.443$ at which the total energy becomes minimum,
whereas the $x$-direction displacements
of Sr and  O$_{\rm II}$
($\equiv y$-direction of O$_{\rm I}$)
become active in the region of $u>0.443$,
as shown in Fig.\ref{fig:TotalEnergyMonoSrTiO3}(c).
While lattice constant $c$
decreases slightly at first,
it changes into the tendency to increase
when the relation of the displacements of $x$-direction
between O$_{\rm I}$ and  O$_{\rm II}$
is reversed.
We analyze the charge density distribution
at each amplitude of atomic displacements $u$ for the tensile strained SrTiO$_3$.
Shown in Fig.\ref{fig:ChargeMonoSrTiO3}
are the pseudo valence charge density maps
in the (001) cross sections and a projection
at the following three points;
[A] paraelectric state at $u=0$,
[B] monoclinic equilibrium state at $u=0.443$ Bohr, and
[C] large displacement state at $u=0.870$ Bohr,
which are printed on total-energy surface
shown in Fig.\ref{fig:TotalEnergyMonoSrTiO3}(a).
In the process
from the paraelectric state to the monoclinic equilibrium state,
we can find that Ti comes close to O$_{\rm I}$ and O$_{\rm II}$
while Sr and O$_{\rm III}$
approach mutually,
shown in Fig.\ref{fig:ChargeMonoSrTiO3}[A],[B].
In other words,
the closenesses of Ti-O$_{\rm I}$,O$_{\rm II}$
and that of Sr-O$_{\rm III}$
play the dominant role
for ferroelectric structural distortion
in SrTiO$_{3}$ induced 4\% tensile strain.
In the region of the large amplitude of atomic displacements $u>0.443$,
however,
the displacements
which O$_{\rm I}$ and O$_{\rm II}$ come close to Sr
are enhanced
while the closenesses of Ti-O$_{\rm I}$,O$_{\rm II}$
and that of Sr-O$_{\rm III}$
are saturated respectively,
as shown in Fig.\ref{fig:TotalEnergyMonoSrTiO3}(c) and Fig.\ref{fig:ChargeMonoSrTiO3}[C].
From these results,
the displacements
of O$_{\rm I}$ and O$_{\rm II}$
are shifted in the direction
in which they approach Sr
because the Ti-O$_{\rm I}$,O$_{\rm II}$ closeness displacement
is saturated by the rigid-sphere-like ionic repulsion between them,
so in the atomic displacements of $x$-direction in $u>0.443$
as shown in Fig.\ref{fig:TotalEnergyMonoSrTiO3}(c),
Sr in the positive charge and
O$_{\rm II}$ in the negative charge
are displaced dominantly.
The increase of lattice constant $c$
works to relax the rapid closeness of Sr-O$_{\rm I}$,O$_{{\rm II}}$.
Hence,
SrTiO$_3$ can be regarded as a typical ionic crystal.
The origin of the ferroelectricity can be considered that
the rigid-sphere-like ions
go into open spaces induced by the biaxial tensile strain.

\section{Summary}
\label{sec:summary}
In this study,
we investigated the effect
of the in-plane biaxial compressive and tensile strains of SrTiO$_{3}$
on the ferroelectric distortion
by determining the total-energy surface accurately
with our structure optimization technique.
(a) We confirmed that
the ferroelectric distortion is induced strongly
and Curie temperature is raised
by applying the biaxial compressive or tensile strains
to SrTiO$_{3}$.
(b) From the results of the normalized atomic displacements,
it is clarified that
the ferroelectric distortion of SrTiO$_{3}$
is strongly influenced by the atomic displacements
corresponding to the hard mode
as well as the soft mode.
(c) By analyzing the obtained normalized displacements
for 4\% biaxial compressive strain,
we showed that
the Slater mode plays a dominant role
while the Last mode is also important
in the ferroelectric phase transition of SrTiO$_{3}$.
(d) The behavior of atomic displacement
associated with ferroelectric distortion
was clarified by analyzing the charge density distribution
in each amplitude of atomic displacements
for 4\% biaxial tensile strain.

%

\section*{Acknowledgements}
Computational resources
were provided by the Center for Computational Materials Science,
Institute for Materials Research (CCMS-IMR), Tohoku University.
We thank the staff at CCMS-IMR for their constant effort.

\bibliographystyle{jjap}
\bibliography{biblio/abinitio,biblio/ferroelectrics,biblio/ChemicalReaction,biblio/Perovskite-likeFluoride,biblio/LandoltBornstein}
\end{document}